 \documentclass[iop]{emulateapj} 

\shorttitle{Hierarchical Solar-Type Multiple Star Systems}
\shortauthors{Roberts et al.}

\begin{document}


\title{Continued Kinematic and Photometric Investigations of Hierarchical Solar-Type Multiple Star Systems}

\author{
Lewis C. Roberts, Jr.\altaffilmark{1}, 
Andrei Tokovinin\altaffilmark{2},
Brian D. Mason\altaffilmark{3}, 
Anne D. Marinan\altaffilmark{1} 
} 

\altaffiltext{1}{Jet Propulsion Laboratory, California Institute of Technology, 4800 Oak Grove Drive, Pasadena CA 91109, USA}
\altaffiltext{2}{Cerro Tololo Inter-American Observatory, Casilla 603, La Serena, Chile}
\altaffiltext{3}{U.S. Naval Observatory, 3450 Massachusetts Avenue, NW, Washington, DC 20392-5420, USA}
\email{lewis.c.roberts@jpl.nasa.gov}
 
\label{firstpage}

\begin{abstract}

We observed 15 of the solar-type binaries within 67 pc of the Sun previously observed by the Robo-AO system in the visible, with the PHARO near-IR camera and the PALM-3000 adaptive optics system on the 5 m Hale telescope.  The physical status of the binaries is confirmed through common proper motion and detection of orbital motion. In the process we detected a new candidate companion to HIP 95309. We also resolved the primary of HIP 110626 into a close binary making that system a triple. These detections increase the completeness of the multiplicity survey of the solar-type stars within 67 pc of the Sun. Combining our observations of HIP 103455 with archival astrometric measurements and RV measurements, we are able to compute the first orbit of HIP~103455  showing that the binary has a 68 yr period. We place the components on a color-magnitude diagram and discuss each multiple system individually.  

\end{abstract}

  

\section{Introduction}\label{sec:intro}

We report observations of stellar multiple systems  with adaptive optics (AO), completing the  program of  \citet[][hereafter Paper I]{roberts2015}. The program's goal is the characterization  of  hierarchical multiplicity of solar-type stars in the solar neighborhood. This is a follow-up of the large  multiplicity survey conducted with the Robo-AO system  \citep[][R15]{riddle2015}. As in Paper I, we observed multiple systems discovered in R15 to confirm the bound nature of their companions by relative  astrometry and infrared (IR) photometry.  New subsystems discovered serendipitously in Paper I were re-visited  here to follow their orbital motion, with the eventual goal of determining orbits and masses. In fact, we are already able to compute one orbit by combining  new and archival data of HIP 103455.  

The motivation for  this program was presented in  Paper I.  Continued interest  in  the  multiplicity  of  solar-type  stars,  already  well characterized, is driven  by the need to clarify  the properties of hierarchical systems  with three or more stars.   Hierarchies open new insights  on  formation  mechanisms  of  binaries  by  their  relative frequency, statistics  of period ratios,  eccentricities, and relative orbit  orientations.   Compared  to  binaries,  hierarchies  are  more difficult to study, requiring combinations of observing techniques, so much work remains before we attain a good  knowledge of hierarchical systems  even  in  the  close  solar neighborhood. In addition, the period ratio between the closer and wider pairs of hierarchical multiples differs by at least half an order of magnitude, contributing to the complications for multiplicity studies. Our observations provide an incremental input to this effort.

Additional interest in the  multiplicity of solar neighbors stems from the fact that many of these stars host exoplanets \citep{raghavan2006,bonavita2007}. The  emerging statistics show  that  exoplanets  are  ubiquitous,  so many  stars  not currently known as exoplanet hosts will join  this family  in the  future.  Planet formation and evolution are closely  related to the multiplicity of the host stars, hence the need  to explore the multiplicity of these stars as best as we can. 

New observations are described in Section \ref{sec:obs}, their results are  given in Section~\ref{sec:res}.   Comments on  individual systems are given in Section~\ref{sec:comments}, while Section~\ref{sec:orbit} presents the combined visual-spectroscopic orbit of HIP~103455. The results are summarized in Section~\ref{sec:summary}.  

\section{OBSERVATIONS}\label{sec:obs}

We observed the  stars on 2015 July 7 UT  with the Palomar Observatory Hale 5 m telescope using the PALM-3000 natural  guide-star AO system \citep{dekany2013} and the PHARO near-IR camera \citep{hayward2001}. We collected 50 frames of each  object with an  exposure time of 1.416 s, which  is the minimum exposure for  PHARO. The data were reduced by debiasing, flat fielding, bad  pixel  correction  and  background subtraction and then  shift-and-added  to  create a  single  image.  The \textit{fitstars}  algorithm was  used to  measure the  astrometry and photometry of the objects \citep{tenBrummelaar1996,tenBrummelaar2000}.

We observed three  calibration binaries\footnote{\url{http://ad.usno.navy.mil/wds/orb6/orb6c.html}} on the same night as  the science targets. See Table \ref{cal_stars}. We compared their measured astrometry with the ephemeris predicted from their orbits and used the results  to compute  the plate  scale and  the position  angle offset, 24.9$\pm$0.2  mas  pixel$^{-1}$  and 0.7$\pm$0.5$^\circ$. Since astrometric calibration based on orbits include the errors in the orbits, we have presented the measured differential pixel  locations and the orbits used to calibrate the binaries in Table \ref{cal_stars}. This allows the reader to judge the calibration accuracy and for future readers to recompute the astrometric calibration if improved orbits become available. Error bars on the astrometry were computed using the technique in \citep{roberts2015}. Photometric  error bars were assigned  using the technique described in  \citet{roberts2005}.

\begin{deluxetable*}{llccl}
\tabletypesize{\scriptsize}
\tablewidth{0pt}
\tablecaption{Calibration Binaries \label{cal_stars}}
\tablehead{\colhead{WDS} &\colhead{Discovery Des.} & \colhead{$\Delta$x (pix.)} & \colhead{$\Delta$y (pix.)} & \colhead{Orbit Reference} }
\startdata
15360$+$3948 & STT  298AB & \phd\phn\phn4.4 &\phn\phd47.3 &  \citet{soderhjelm1999}\\
16147$+$3352 & STF 2032AB & \phd246.8       &\phd146.5    &  \citet{raghavan2009}\\
16160$+$0721 & STF 2026AB & \phn-42.7       &-134.0       &  \citet{scardia2011} 
\enddata
\end{deluxetable*}

\section{Astrometry, Photometry, and CMDs}\label{sec:res}

New information resulting from our observations is highlighted in this Section. Table~\ref{results} contains information on 15 solar-type systems: 4 binaries, 9 triples,  1 quadruple, and 1 quintuple.  We are able to confirm the physical nature of four Robo-AO candidate companions: HIP~94666, HIP~94905, HIP~95309, and HIP~110626. The table gives the Washington Double Star (WDS) Catalog designation for the system as well as the HD  and HIP numbers for the primary star.  This is followed by the discoverer designation, \textbf{the known mulitplicity of the system,} the  observation epoch, the astrometry, and the differential photometry of the pair listed in the discovery designation.  \textbf{For systems where we detected more than two components in the system, we have only listed the multiplicity for the first pair.}  New discoveries are marked with a footnote to the WDS designation. For some of the closer systems we used an alternative blind-deconvolution code and these are also marked with a footnote.  

\begin{deluxetable*}{lrrl c cccccc}
\tabletypesize{\scriptsize}
\tablewidth{0pt}
\tablecaption{Relative Astrometry and Photometry\label{results}}
\tablehead{\colhead{WDS} &\colhead{HD}&\colhead{HIP}&\colhead{Discoverer}& \colhead{Mult} &\colhead{Epoch}&\colhead{$\theta$ (\degr)} & \colhead{$\rho$ (\arcsec) } & \colhead{$\Delta$J}& \colhead{$\Delta$H} & \colhead{$\Delta$Ks}}
\startdata
14094$+$1015\tablenotemark{a} & 123760 & 69160 & RAO 16        &2& 2015.5141  &  249.8$\pm$1.7  &   0.11$\pm$0.01    & 0.25$\pm$0.1\phn     & - &  0.27$\pm$0.1\phn\\ 
15277$+$4253 & 138004 &  75676 & RAO 18 BC                     &5& 2015.5136  &  112.8$\pm$1.6  &   0.37$\pm$0.01  & 1.10$\pm$0.10 & - & 0.99$\pm$0.04\\
17229$-$0223 & 157347 &  85042 & RAO 19 BC                     &3& 2015.5138  &  \phn89.8$\pm$0.9  &   0.73$\pm$0.01  & 0.01$\pm$0.01 & - & 0.04$\pm$0.01\\
17422$+$3804\tablenotemark{a} & 161163 &  86642 & RBR 29 Aa,Ab &3& 2015.5138  &  \phn35.0$\pm$7.8  &   0.07$\pm$0.01  & - & - & 1.44$\pm$0.1\phn\\  
17422$+$3804\tablenotemark{a} & 161163 &  86642 & RAO 20 Aa,B  & & 2015.5138  &  303.3$\pm$0.6  &   2.25$\pm$0.01   & - & - & 3.60$\pm$0.1\phn\\
18352$+$4135\tablenotemark{a} & 171886 &  91120 & RAO 83 Ba,Bb &3& 2015.5137  &  \phn16.0$\pm$7.2  &   0.08$\pm$0.01   & 0.96$\pm$0.1\phn & - & 1.11$\pm$0.1\phn\\      
19158$+$3823 & 180683 &  94666 & RAO 85                        &3& 2015.5139  &  331.8$\pm$0.5  &   3.58$\pm$0.03  & 5.05$\pm$0.17 & - & 4.45$\pm$0.04\\  
19188$+$1629 & 181144 &  94905 & RAO 67                        &3& 2015.5141  &  190.5$\pm$0.5  &   6.86$\pm$0.03  & 4.18$\pm$0.07 & - & 3.45$\pm$0.03\\  
19234$+$2034\tablenotemark{b} & 182335 &  95309 & NEW AC       &3& 2015.5140  &  \phn\phn8.1$\pm$0.5  &   3.06$\pm$0.02  & 3.80$\pm$0.07 & - & 2.25$\pm$0.03 \\ 
19234$+$2034 & 182335 &  95309 & RAO 68 AB                     & & 2015.5140  &  351.0$\pm$0.5  &   5.07$\pm$0.02  & 3.95$\pm$0.07 & - &3.43$\pm$0.03\\  
20312$+$5653\tablenotemark{a} & 195872 & 101234 & RAO 22       &2& 2015.5142  &  151.6$\pm$3.4  &   0.17$\pm$0.01  & 2.05$\pm$0.43 & - & 1.9\phn$\pm$0.1\phn\\ 
20333$+$3323\tablenotemark{a}& 195992 & 101430 &RBR 29 Ba,Bb   &4& 2015.5140  &  299.1$\pm$2.9  &   0.20$\pm$0.01   & -     & -   & 0.33$\pm$0.1\phn\\
20577$+$2624 & 199598 & 103455 & RAO 24                        &2& 2001.4873  &  332.0$\pm$2.8    &   0.24$\pm$0.01    & -  & 2.9$\pm$0.19\tablenotemark{c} & -\\   
             &        &        &                               & & 2015.5142  &  \phn93.7$\pm$0.9  &   0.72$\pm$0.01  & 3.23$\pm$0.26  & - & 2.74$\pm$0.08\\   
22094$+$3508 & 210388 & 109361 & RAO 26                        &2& 2015.5143  &  \phn\phn4.5$\pm$1.2 &    0.41$\pm$0.01  & 2.21$\pm$0.19 & & 1.76$\pm$0.04\\   
22246$+$3926 & 212585 & 110626 & RAO 28 AaB                    &3& 2015.5143  &  316.1$\pm$0.5 &    4.42$\pm$0.01 & 4.83$\pm$0.17 & 4.47$\pm$0.07 & 4.13$\pm$0.03\\
22246$+$3926\tablenotemark{a,b} & 212585 & 110626 & NEW Aa,Ab  & & 2015.5143  &  262.2$\pm$1.7 &    0.22$\pm$0.01 & 3.38$\pm$0.1\phn  & 3.12$\pm$0.1\phn & 3.08$\pm$0.1\phn\\
22266$+$0424\tablenotemark{a}   & 212754 & 110785 & BU 290 AB  &3& 2015.5141  &  226.2$\pm$0.8 &    3.90$\pm$0.02 & -   & - &     3.68$\pm$0.1\phn\\
23588$+$3156 & 224531 & 118213 & RBR 30 Aa,Ab                  &3& 2015.5144  &  351.2$\pm$1.4 &    0.43$\pm$0.01  &- &- & 4.36$\pm$0.26\\   
23588$+$3156 & 224531 & 118213 & RAO 76 Aa,B                   & & 2015.5144  &  \phn87.5$\pm$0.5 &    4.85$\pm$0.02  &- &- & 4.84$\pm$0.04\\  
\enddata
\tablenotetext{a}{Data were processed with alternative blind-deconvolution method.}
\tablenotetext{b}{New Discovery}
\tablenotetext{c}{Data was taken with an FeII filter centered at 1.65 $\mu$m.}
\end{deluxetable*}

\begin{figure}[ht]
   \centering
   \includegraphics[width=8cm]{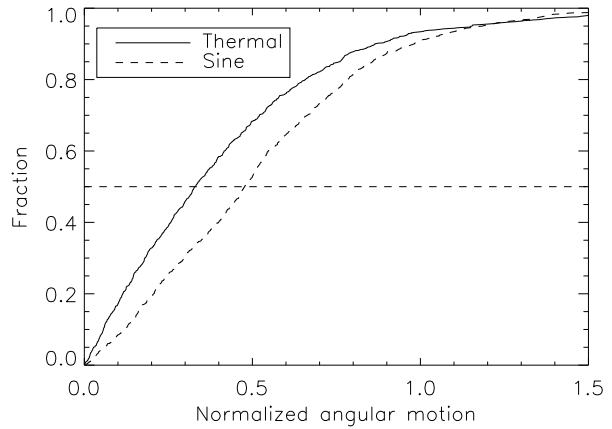} 
   \caption{The cumulative distributions of the angular speed normalized by the circular-orbit speed 360\degr $P^{-1}$.  Randomly oriented orbits with eccentricities distributed as $\sim\sin(\pi e)$ (sine) and $2e$  (thermal) are simulated. 
\label{simulation}}
\end{figure}

For the observed systems, we have estimated their orbital periods. There are a few ways to estimate the periods of binaries with only a handful of astrometric measurements. One, $P^*$, is based on the projected separation when we know the distance to the system \citep{FG67a}; in most cases it is within a factor of three from the true period.  The other is based on the angular velocity and the assumption that the system has a circular face-on orbit.  Both techniques provide rough orders of magnitude estimates, but which is more accurate? To help answer this question, we simulated a population of binaries with random orientation and with the  thermal and sine eccentricity distributions \citep[][average eccentricity 0.67 and 0.5, respectively]{tokovinin2016}. Figure \ref{simulation} shows the cumulative distribution of the angular speed normalized by the circular-orbit speed. The medians are 0.33 and 0.48 for the thermal and sine eccentricity distributions.     So, binary period estimated from angular speed, $P_t = 360\degr \dot{\theta}^{-1}$, where $\dot{\theta}$ is the rate of position angle change, is longer than the true period by a factor of 2 to 3 in half of the cases and longer than the true period in 90\% of the cases. Binaries spend a significant fraction of time moving slower than average and ``catch up'' to complete the full revolution in $P$ years during short time intervals.  Both estimation techniques make only order-of-magnitude, statistical estimates, but $P^*$ is biased less than $P_t$. As such, we provide period estimates based on $P^*$.  
  
\begin{figure*}[ht]
    \centering
\centerline{
\includegraphics[width=8.5cm]{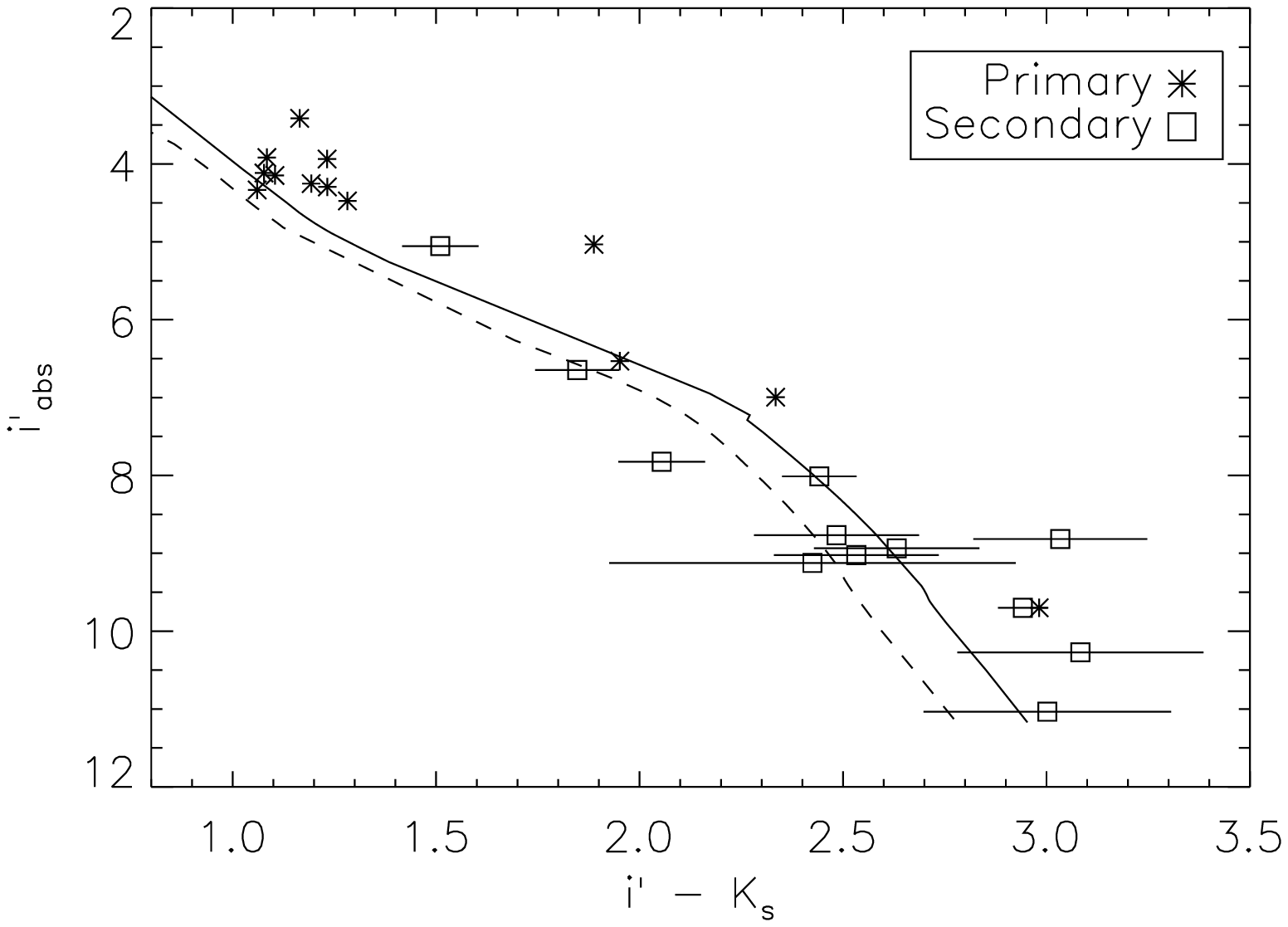}
\includegraphics[width=8.5cm]{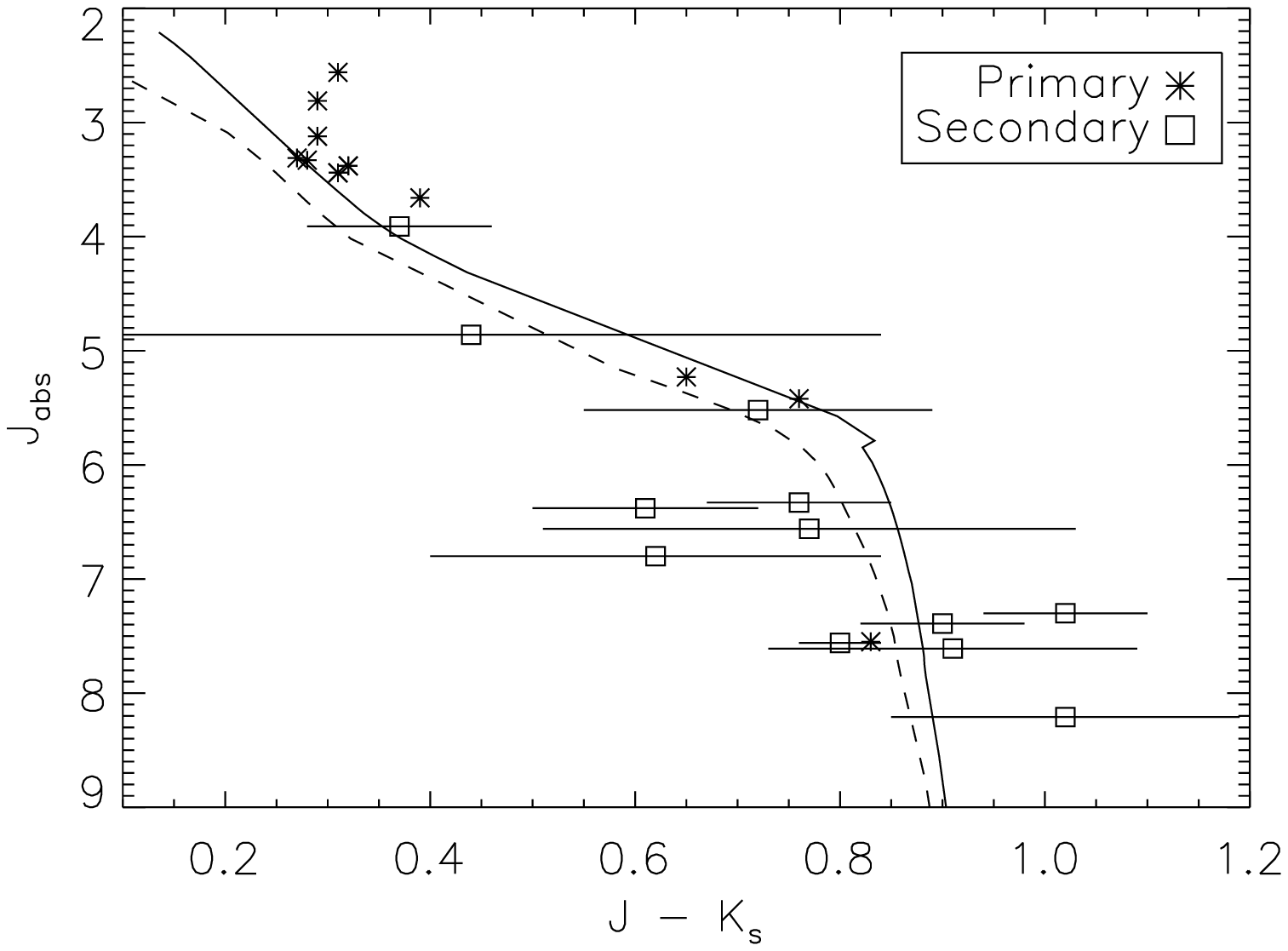}
}
\caption{\label{fig:cmd} Color-magnitude  diagrams.  Left: in  the $i',
  i'-K_s$ bands, right: in the $J, J-K_s$ bands.  Asterisks denote primary
  components, squares with error  bars are secondary components.  Full
  and dashed  lines are Dartmouth  1-Gyr isochrones \citep{Dotter2008}
  for solar metallicity and [Fe/H]=$-$0.5, respectively. }
\end{figure*}

Differential   photometry   in   the   $J$  and   $K_s$   bands   from Table~\ref{results}  and combined magnitudes  of stellar  systems from 2MASS  allow  us to  compute  the  individual  magnitudes of  resolved components and  place them on the color-magnitude  diagram (CMD) using {\it Hipparcos} parallaxes. To extend  the color base, we also use the SDSS $i'$  magnitudes, with the  differential photometry from  R15 and the combined  magnitudes evaluated by interpolation  from other bands, as   explained   in   Paper   I.    The  two   CMDs   are   shown   in Figure~\ref{fig:cmd}.  In  calculating  the  error bars,  we  assigned errors of 0.1 mag to  the differential photometry from R15 and assumed reasonable errors for the combined magnitudes. \textbf{At  an average  distance  of 50pc,  the  a 1-mas parallax  error (the largest of any of our systems) translates  to a  5\% distance  error or  0.1  mag in absolute magnitudes, which is well within the natural CMD uncertainty caused by age and  metallicity differences and imperfect isochrones.} Table~\ref{tab:ptm} lists absolute magnitudes and colors of individual components computed from the differential and combined photometry and used to plot the CMDs in Figure~\ref{fig:cmd}. Using absolute magnitudes, we estimate  masses of secondary components by assuming that they are located on the main sequence. 

 \begin{deluxetable*}{r c c cc cc}
\tabletypesize{\scriptsize}
\tablewidth{0pt}
\tablecaption{Photometry of components \label{tab:ptm}}
\tablehead{
\colhead{HIP} &
\colhead{$\pi_{\rm HIP2}$} &
\colhead{Comp} &
\colhead{$i'_{\rm abs}$} &
\colhead{$J_{\rm abs}$} & 
\colhead{$i'-K_s$} &
\colhead{$J-K_s$} \\
 &  \colhead{(mas)} & &
\colhead{(mag)} &
\colhead{(mag)} &
\colhead{(mag)} &
\colhead{(mag)} 
}
\startdata
 69160  &   17.6$\pm$0.9  & A & \phn4.33 & 3.66  &  1.06$\pm$0.07&    0.39$\pm$0.07\\
        &             & B & \phn5.05 & 3.91  &  1.51$\pm$0.09&    0.37$\pm$0.09\\
 75676  &   31.1$\pm$0.6 & B & \phn6.53 & 5.23  &  1.95$\pm$0.04&    0.65$\pm$0.05\\
        &             & C & \phn8.01 & 6.33  &  2.44$\pm$0.09&    0.76$\pm$0.09\\
 85042  &   51.2$\pm$0.4 & B & \phn9.70 & 7.55  &  2.98$\pm$0.06&    0.83$\pm$0.04\\
        &             & C & \phn9.70 & 7.56  &  2.94$\pm$0.06&    0.80$\pm$0.04\\
 86642  &   24.7$\pm$0.5 & A & \phn3.94 & 3.94  &  1.23$\pm$0.04&    1.23$\pm$0.04\\
        &             & B & \phn8.94 & 8.94  &  2.63$\pm$0.20&    2.63$\pm$0.20\\
 91120  &   20.2$\pm$0.4 & Ba& \phn6.99 & 5.42  &  2.33$\pm$0.06&    0.76$\pm$0.05\\
        &             & Bb& \phn7.82 & 6.38  &  2.05$\pm$0.11&    0.61$\pm$0.11\\
 94666  &   15.5$\pm$0.6 & A & \phn3.41 & 2.56  &  1.16$\pm$0.04&    0.31$\pm$0.04\\
        &             & B & \phn9.12 & 7.61  &  2.42$\pm$0.50&    0.91$\pm$0.18\\
 94905  &   26.7$\pm$0.6 & A & \phn3.92 & 3.12  &  1.08$\pm$0.04&    0.29$\pm$0.04\\
        &             & B & \phn8.77 & 7.30  &  2.48$\pm$0.20&    1.02$\pm$0.08\\
 95309  &   21.7$\pm$1.0 & A & \phn4.29 & 3.44  &  1.23$\pm$0.04&    0.38$\pm$0.04\\
        &             & B & \phn9.02 & 7.39  &  2.53$\pm$0.20&    0.90$\pm$0.08\\
        &             & C &  \ldots  & 7.24  &     \ldots        &    1.85$\pm$0.08\\ 
101234  &   15.1$\pm$0.9 & A &  \ldots  & 2.81  &     \ldots        &    0.29$\pm$0.08\\
        &             & B &  \ldots  & 4.86  &     \ldots        &    0.44$\pm$0.40\\
103455  &   31.6$\pm$0.6 & A & \phn4.15 & 3.33  &  1.10$\pm$0.04&    0.28$\pm$0.04\\
        &             & B & \phn8.82 & 6.56  &  3.03$\pm$0.21&    0.77$\pm$0.26\\
109361  &   23.3$\pm$0.6 & A & \ldots   & 3.31  &     \ldots        &    0.27$\pm$0.04\\
        &             & B & \ldots   & 5.52  &     \ldots        &    0.72$\pm$0.17\\
110626  &   20.3$\pm$0.9 & A & \phn4.25 & 3.31  &  1.19$\pm$0.04&    0.31$\pm$0.04\\
        &             & B &10.27 & 5.52  &  3.08$\pm$0.30&    1.02$\pm$0.17\\
        &             & C &  \ldots  & 6.80  &       \ldots      &    0.62$\pm$0.22\\
118213  &   20.2$\pm$0.7 & A &  \phn4.48& \ldots    &  1.28$\pm$0.04&        \ldots       \\
        &             & B & 11.04& \ldots    &  3.00$\pm$0.30&        \ldots       
\enddata
\end{deluxetable*}
 
\section{Comments on individual systems}\label{sec:comments}

{\it HIP  69160 (HD 123760 = WDS 14094+1015)} is an SB2 with an 8-yr period.  The SB2 designation for this pair comes from \citep{FG67a},  which also provided the input for the R15 survey observing list, which Paper I and this work are now following up.  A companion was resolved by R15 and it is confirmed by our observations. The companion's position angle has changed by 11\degr~in the 2.5 years since it was discovered.  The semi-major axis estimated from  period is 92\,mas, which is close to the measured separation.  It seems likely that the system is near apastron and is at its slowest part of the orbit. If so the eccentricity of the orbit is probably high.  The alternative that this is a wider tertiary companion to an unresolved SB2 is unlikely, considering that small $\Delta m$ matches the large spectroscopic mass ratio and the separation matches the SB2 axis.

\textbf{{\it HIP 75676 (HD 138004 = WDS 15277+4253)} (A), together with BC at 40\farcs2, form a wide physical pair.}  The BC binary was first observed by R15 and confirmed by our observation. Its component B is a 17-day spectroscopic pair \citep{FG67a}, while the star A is a spectroscopic  and astrometric binary \citep{MK05}, meaning that this is a quintuple stellar system.  We provide  IR photometry of the pair BC with an estimated period of 40 yr. Its position angle changed by 13\degr.  While the magnitude of the proper motion motion and the orbital motion are similar, they are moving in different directions and we conclude that this system has common proper motion. 

{\it HIP 85042 (HD 157347  = WDS 17229-0223)} is a visual triple, with a wide 46\arcsec AB pair \citep{raghavan2010} and the BC pair consisting of two M dwarfs (R15). Our observations confirm the earlier detection of R15. The  estimated period of BC is $\sim$40 yr.  The magnitude of the proper motion of the primary is roughly 10 times the orbital motion, leading us to conclude this system has common proper motion.  

{\it HIP 86642 (HD 161163 = WDS 17422+3804)} is a  triple system.  The 6-yr spectroscopic subsystem Aa,Ab \citep{FG67a} was resolved  in Paper I, where the physical  nature of the wide pair detected by R15  was  also   established. As this system has a short period, it should be observed as often as possible, so that a combined visual/spectroscopic orbit can be computed. 

{\it HIP 91120 (HD 171886 = WDS  18352+4135)} We resolved the subsystem Ba,Bb discovered with Robo-AO at 0\farcs14 and  336\degr, now closing to 0\farcs08  and 16\degr. Its estimated period is 16 yr, the observed orbital motion is indeed fast. The primary's proper motion is larger than the orbital motion including the large error bar on the position angle and we conclude the system has common proper motion. 

{\it HIP 94666 (HD 180683 = WDS 19158+3823)} The physical nature of the 3\farcs6 tertiary companion  to this spectroscopic binary \citep{FG67a} is confirmed by the large proper motion of the primary compared to the tiny orbital motion between the observation of R15 and our current measurement. Our photometry places the secondary on the main sequence, its mass is  $\sim$0.37 ${\cal M}_\odot$.

{\it HIP 94905 (HD 181144 = WDS 19188+1629)}  The nature  of the tertiary companion B, at 6\farcs9 from the spectroscopic binary \citep{FG67a},  was controversial considering that the system is located in the Galactic plane with dense stellar background, while its HIP2 proper motion (PM) is slow,  30\,mas yr$^{-1}$. Comparison of our  astrometry with 2MASS, as well  as our resolved photometry, indicate  that the pair AB is physical. Its position would  have changed by 0\farcs45 since 2MASS epoch if B had a zero PM.

\begin{figure*}[tb]
 \centering
 \begin{tabular}{ccc} 
\includegraphics[height=8cm]{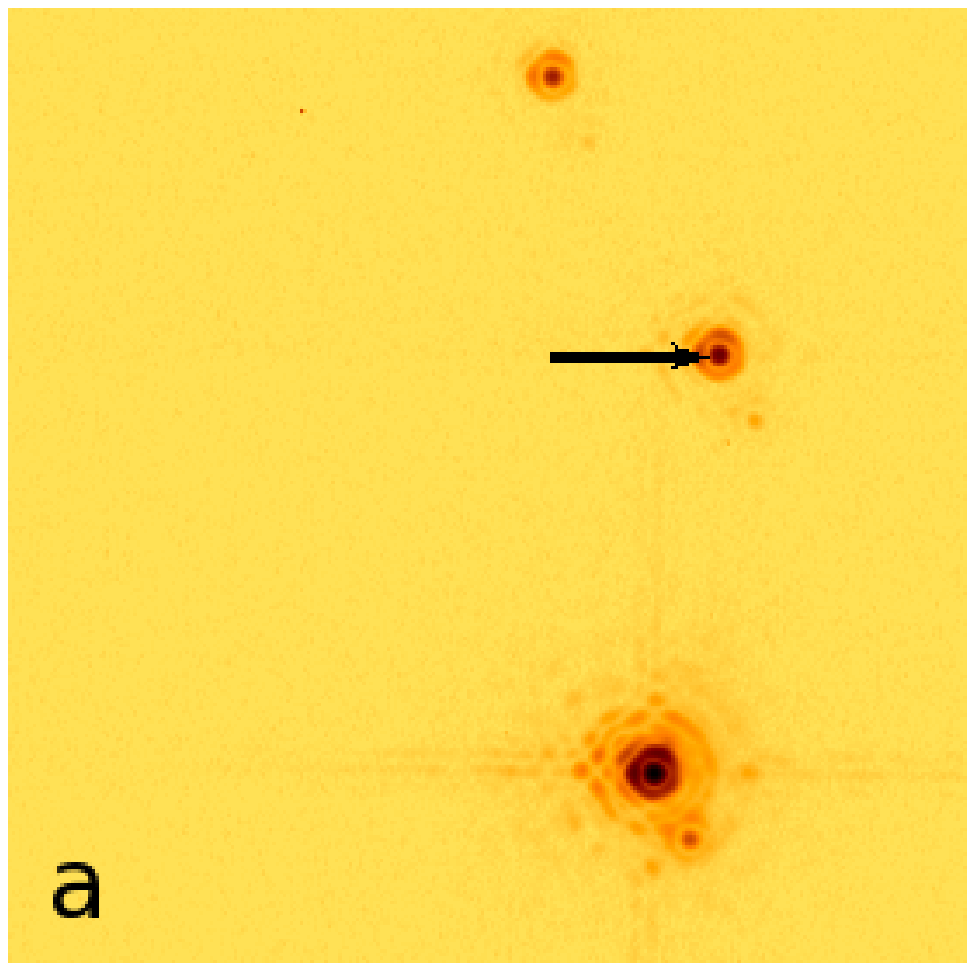} 
\includegraphics[height=8cm]{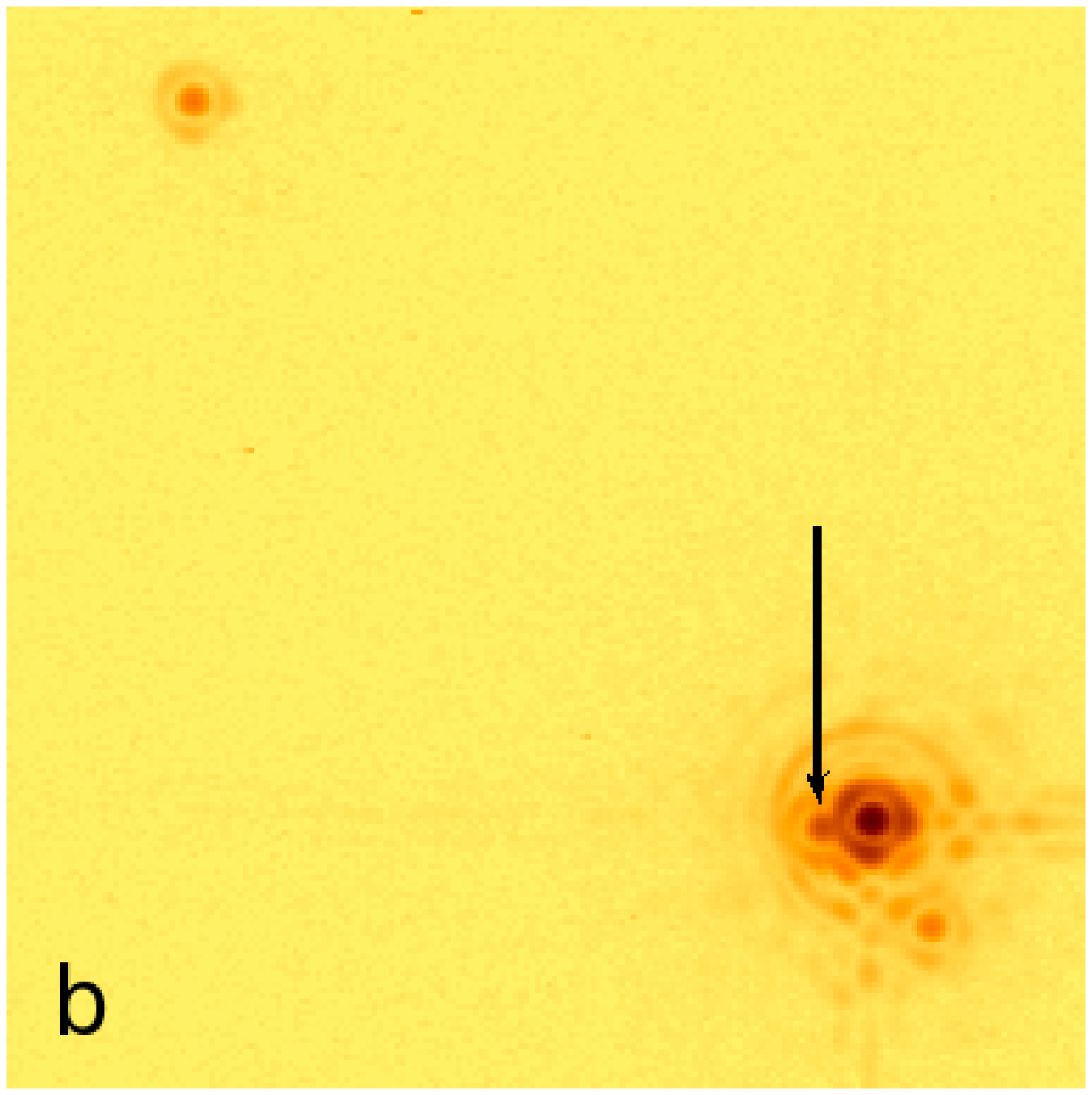} 
 \end{tabular}
 \caption{Images of the  two new components in the $K_s$ band. North is up and East is to the right. The images are: (a) HIP  95309 and (b) HIP 110626. Black arrows point to the locations of the newly imaged companions. The arrows  have lengths of approximately one arcsecond. In each of the images, there is a ghost to the lower right of each star caused by a neutral density filter in the PHARO  camera. These are subimages from the actual data frames; the field of view varies and was chosen to best display the binary. The images are all intensity stretched to best display the binary and  the PSF
   structure. \label{new_binaries} }
 \label{images}%
\end{figure*}

{\it  HIP 95309 (HD 182335 = WDS 19234+2034)}  is  another object  in  the Galactic  plane, with  a moderate PM  of $(+13,  -186)$\,mas yr$^{-1}$.  The physical  nature of  B is confirmed by its stable position and its location on the main sequence in the  CMDs. Unexpectedly, we have  discovered a new  component C with a separation of 3\farcs1, between A and B (see Figure~\ref{new_binaries}(a).) The new component, C,  is brighter than B,  but it was not seen  in the $i'$ band  with Robo-AO (R15).  We  measure its  color $J-K_s  = 1.85$  mag, much redder than any  normal dwarf star.  The nature of  C is enigmatic, as most stars around HIP~95309 in  the 2MASS $K_s$-band image are fainter and it is difficult  to figure out how C could project  on the AB pair without  being noticed  before.   It  could be  either  a distant  red transient,  or  a  nearby  low-mass  brown  dwarf  with  a  very  fast PM. 

{\it HIP 101234 (HD 195872 = WDS 20312+5653)} was  re-observed to confirm its substantial magnitude difference established in Paper I  and discrepant with the $\Delta i' \sim 0$ measured  with Robo-AO.  The  period is about  30 yr, and  the pair turned  by 44\degr~since  2012.76. The  observed  motion matches  the estimated  period,  but more  data  are  needed  before attempting  to compute the orbit.  The differential magnitudes measured in 2015 are within the error bars of the differential magnitudes measured in Paper I and do not shed any light on why they are so different than those measured in R15. 

{\it HIP 101430 (HD 195992 = WDS 20333+3323)} was observed in Paper I and the B component was revealed to be a close binary with an estimated period of $\sim$30 years.  This paper's observations show direct motion and the binary opening up.  The A component was saturated and we were unable to measure the astrometry of the A,Ba or AE pairs.

{\it HIP 103455 (HD 199598 = WDS 20577+2624)} The new measurement combined with existing data leads to the determination of the orbit in Section \ref{sec:orbit}.

{\it HIP 109361 (HD 210388 = WDS 22094+3508)} is a  18.5-yr spectroscopic binary \citep{FG67a} that turned by $13^\circ$ since its resolution with Robo-AO. 
 
{\it HIP 110626 (HD 212585 = WDS 22246+3926)}  We resolved the inner astrometric \citep{MK05}and spectroscopic subsystem Aa,Ab  \citep{FG67a} at 0\farcs22  separation  (see Figure~\ref{new_binaries}(b)).  From  the projected separation, the estimated orbital period is $\sim$30yr.  Robo-AO was unable to resolve the system due to the large contrast, $\Delta K=3.1$, between the two components. The component B has the luminosity of a $\sim$0.23\,${\cal M}_\odot$ dwarf.

{\it HIP 110785 (HD 212754 = WDS 22266+0424)} is a triple system with a wide pair in a 420 year period \citep{hale1994}, with a primary consisting of a spectroscopic binary \citep{FG67a} with a 931 day period \citep{griffin2010}.  Despite  the spectroscopic binary's semi-major axis  of 59\,mas, we were unable to resolve the subsystem  owing  to the small mass ratio  of 0.22 determined  from the spectroscopic and astrometric orbits \citep{goldin2007}. 
 
{\it HIP 118213 (HD 224531 = WDS 23588+3156)}  is a triple system similar  to HIP~110626, where the inner astrometric \citep{MK05} and spectroscopic pair Aa,Ab \citep{FG67a} was first  resolved in Paper I.   We confirm  the large  magnitude  difference between  Aa and  Ab, $\Delta K_s  = 4.4$ mag. The  separation implies an  orbital period of $\sim$100\,yr;    the   mass  of  Ab  is  estimated   at  0.2  ${\cal   M}_\odot$. The AB pair shows hardly any motion between the R15 observation in 2012 and 2015. The proper motion of the system is high enough that the system is clearly physical.

\section{Orbit of HIP 103455}\label{sec:orbit}

The solar-type  star HIP~103455 (HD 199598, G0V,  $V=6.90$) has a distance of 30\,pc \citep{vanLeeuwen2007}.  It is an  X-ray source  1RXS J205739.4+262420, indicating  relative youth \citep{Guillout}.  It was observed  with AO at Palomar in 2003  among  other  young  stars and  found  unresolved  \citep{MH09}. However, the Robo-AO system resolved it into  a 0\farcs6 binary in 2012  and 2013 \citep{riddle2015}.   This was  confirmed in  July 2015 (Paper I). The separation  is increasing, showing some orbital motion, with an  estimated period of $\sim$60\,yr.   S.~Metchev (2015, private communication) re-examined  the Palomar data of 2003  and detected the companion overlapping with the first diffraction ring at a separation of 0\farcs1.

The  binarity  of  HIP~103455  was established  independently  by  its variable  radial  velocity   (RV).   Using  precise  RV  measurements, \citet{Patel2007} suggested a probable orbit with a poorly constrained period around  60 yr and  a large eccentricity of approximately 0.7.   The RV amplitude  was small,  only 1.1  km~s$^{-1}$, indicating  a companion  of low mass, $M_{2  \rm min} =  0.12$ ${\cal M}_\odot$. Direct  resolution of this binary means  that the companion is in fact  more massive and the small  RV  amplitude  is   caused  by  the  low  orbital  inclination. Interestingly, \citet{nidever2002}  did not detect  any significant RV variation over one year in the initial RV data from Lick.

The third independent evidence  of binarity comes from the astrometry. The PM differs from the long-term ground-based PM by 7\,  mas~yr$^{-1}$. This is a  so-called $\Delta \mu$  binary according to \citet{MK05}.  Modeling  detection of such  astrometric binaries shows that their typical periods are a few decades \citep{Tok11}.

We searched the data archives and located images  of HIP~103455 taken with the Gemini Telescope and the Hokupa\`{}a AO  system using the QUIRC science camera \citep{graves1998}. In the images the star was located at different positions across the detector.  We fitted a plane to each image and subtracted it off.  Next, we subtracted off the median  of all the ``de-planed'' images.   This resulted in 22 images.   We  then  measured   the  astrometry  of  each  image  using \textit{fitstars}.  The  error  bars  were set equal  to the  standard deviation   of  the  measurements.     

We computed a combined orbital solution using the ORBITX\footnote{\url{http://www.ctio.noao.edu/$\sim$atokovin/orbit/index.html}} code \citep{tokovinin1992}. ORBITX uses the Levenberg-Marquardt method to solve for all the orbital elements.  The  resolved  measurements  come  from the archival QUIRC data,  S.~Metchev  (2015, private  communication), Robo-AO  \citep{riddle2015},  PHARO \citep{roberts2015}, and  this paper. The individual  RVs were published by  \citet{Fischer14}, covering the period from 1998.5  to 2011.8. The RV measurement accuracy is about 10 m~s$^{-1}$.   Passage through the  periastron in this eccentric orbit has been well  observed. 

\begin{figure}[ht]
    \centering
\includegraphics[width=8cm]{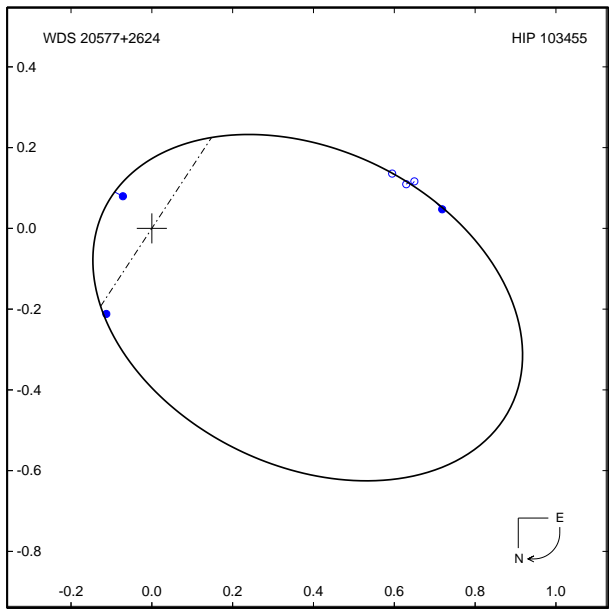}
\caption{Preliminary orbit of HIP~103455.  The broken line through the origin is the line of nodes. Open circles are previously published astrometry values and the filled circles are the new values from this paper. The axes are labeled in units of arcseconds. 
\label{fig:orb_POS}}
\end{figure}

\begin{figure}[ht]
    \centering
\includegraphics[width=8cm]{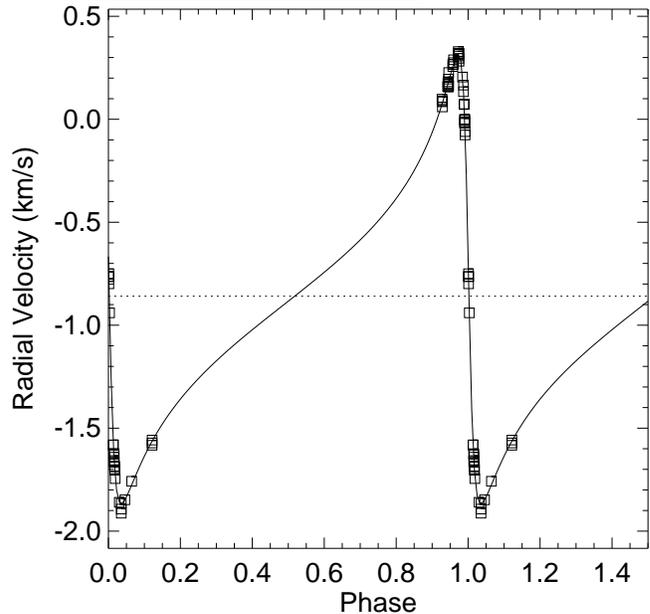}
\caption{The RV orbit of of HIP~103455. The black line is the computed orbit, while the individual data points are overplotted.\label{fig:orb_RV}}
\end{figure}

The orbit is shown in Figures~\ref{fig:orb_POS} and \ref{fig:orb_RV}, and its elements are given in Table \ref{orbital_elements}. Astrometric measurements and their deviations from the orbit are assembled in Table~\ref{tab:measures}.  Altogether, the orbit matches the available data quite  well, with weighted rms residuals of 0\fdg5 in angle, 10\,mas in separation, and 12 m~s$^{-1}$ in RV.  The mass sum is 1.62 ${\cal   M}_\odot$. The RV amplitude  and inclination give the companion mass $M_2$ of 0.27 ${\cal M}_\odot$,  assuming 1.1 ${\cal M}_\odot$ for the primary.  However,  the  large  error  of  inclination  implies  large uncertainty of $M_2$. 

\begin{deluxetable}{ll}
\tabletypesize{\footnotesize}
\tablewidth{0pt}
\tablecaption{Orbital elements of HIP 103455 \label{orbital_elements}}
\tablehead{\colhead{Element}      & \colhead{Value}}
\startdata
$P$ (yr)       &  68.2$\pm$1.5\\
$a$ (\arcsec)  &  0.620$\pm$0.025\\
$i$ (\degr)    &  154.3$\pm$8.5\\
$\Omega$ (\degr)    &  326.7$\pm$2.0\\
$T$            &  2003.4771$\pm$0.0006\\
$e$            & 0.774$\pm$0.004\\
$\omega$ (\degr)    & 84.3$\pm$0.2\\
$K_1$ (kms$^{-1}$) & 1.091$\pm$0.001\\
$V_0$ (kms$^{-1}$) & $-$0.859$\pm$0.002
\enddata
\end{deluxetable}

\begin{deluxetable*}{rrrrccl} 
\tabletypesize{\footnotesize}
\tablewidth{0pt}
\tablecaption{Astrometry and residuals of HIP 103455 \label{tab:measures}}
\tablehead{
\colhead{Date}  &\colhead{$\theta$}       &\colhead{$\rho$}     &\colhead{$\sigma$} &\colhead{(O$-$C)$_\theta$} & \colhead{(O$-$C)$_\rho$} & \colhead{Instrument/} \\
\colhead{(yr)} &\colhead{(\degr)} &\colhead{(\arcsec)} &\colhead{(\arcsec)}    &\colhead{(\degr)}    &\colhead{(\arcsec)} & \colhead{Reference}
}
\startdata
  2001.487 &  332.0 &  0.238 &  0.020 & \phs2.9 &$-$0.001 &  QUIRC/Gemini 8m\\
  2003.720 &  228.0 &  0.110 &  0.010 & \phs2.3 &$-$0.019 & Metchev (Private Comm.)\\
  2012.766 &  102.8 &  0.610 &  0.010 & \phs0.1 & $-$0.004 & \citet{riddle2015}\\
  2013.622 &   99.8 &  0.639 &  0.010 & $-$0.4 & $-$0.005 & \citet{riddle2015}\\
  2013.837 &  100.1 &  0.660 &  0.010 & \phs0.5 & \phs0.009 & \citet{roberts2015}\\
  2015.514 &   93.7 &  0.72\phn  &  0.010 & \phs0.9 &\phs0.015 & PHARO/Hale 5m 
\enddata
\end{deluxetable*}

The period  and mass ratio of HIP~103455 place it  amongst other common solar-type  binaries.   The  available  precise RV  data  exclude  any short-period companions to the main component A, except maybe low-mass planets.  However, the  large eccentricity truncated the circumstellar disk  and likely  prevented formation  of  planets.  The  mass of  the secondary  component B  is constrained  by  its RV  amplitude and  its brightness.   It is  located  above  the main  sequence  in the  $(i', i'-K_s)$  CMD and  on the  main sequence  in the  $(J, J-K_s)$  CMD in Figure~\ref{fig:cmd}, while A lies  on the main sequence. The Galactic velocity  $(U,V,W) =  (-44.4,  -16.4, -16.8)$  km~s$^{-1}$, computed  without correction  for   orbital  motion,  corresponds  to   the  young  disk kinematics  but does  not match  any  known kinematic  groups. If  the component  B is  indeed  a  pre-main-sequence star,  its  mass may  be measured  accurately   in  the   future  by  monitoring   the  orbital motion. The separation is large enough that it is possible to collect individual spectra under good seeing  (or with AO). These spectra can then be used to derive stellar parameters and look for signatures of young age. 
 
\section{Summary}\label{sec:summary}

The observations presented in Paper~I resolved for the first time six new pairs,  some of  them totally unexpected.   Here we add  the first resolution of  the spectroscopic  and astrometric pair  HIP~110626 and the  discovery  of  an  enigmatic  red  companion  to  HIP~95309.   We confirmed several pairs from Paper I and started to observe their orbital motion. The physical nature of  several binaries from R15 with separations of several arcseconds is  established using photometry and astrometry presented here.   We derive the first  orbit of the 68-yr  pair HIP~103455 Although it is not yet fully covered by resolved observations and RVs, the  available data  already constrain  the  orbital elements. This system is likely young, and its further study may be of interest for calibrating masses. 

This  work is a  modest and  incremental contribution  to our knowledge of  the  multiplicity  of solar-type  stars.   Eventually planets will be  discovered around some stars observed  here, and then our observations will become even more valuable, adding a ``look-back'' perspective to future studies of those systems.  This aspect is nicely illustrated here by the role  of archival data in the orbital analysis of HIP~103455.


\section{Acknowledgements}

We thank S. Metchev for reprocessing archival data of HIP 103455. We thank  the staff  of the Palomar  Observatory for  their invaluable assistance  in   collecting  these  data.  This  paper  contains observations    obtained    at    the    Hale    Telescope,    Palomar Observatory. It is based in part on observations  obtained at the Gemini Observatory acquired through the Gemini Science  Archive, which is operated by the Association of  Universities for Research in Astronomy,  Inc., under a cooperative  agreement   with  the  NSF   on  behalf  of   the  Gemini partnership:  the  National Science  Foundation  (United States),  the National  Research Council (Canada),  CONICYT (Chile),  the Australian Research   Council   (Australia),   Minist\'{e}rio   da   Ci\^{e}ncia, Tecnologia  e  Inova\c{c}\~{a}o (Brazil)  and  Ministerio de Ciencia, Tecnolog\'{i}a e Innovaci\'{o}n Productiva (Argentina).  This paper is based  on  observations  obtained  with  the  Adaptive  Optics  System Hokupa\`{}a/Quirc,  developed and  operated by  the University  of Hawaii Adaptive  Optics  Group,  with   support  from  the  National  Science Foundation. A portion of the research in this paper was carried out at the  Jet Propulsion  Laboratory, California  Institute  of Technology, under   a   contract  with   the   National   Aeronautics  and   Space Administration (NASA).
A.M. was supported by a NASA Space Technology Research Fellowship.  This research made use of the Washington Double Star  Catalog maintained  at the  U.S. Naval  Observatory,  the SIMBAD database,  operated  by  the  CDS  in Strasbourg,  France  and  NASA's Astrophysics Data  System. This publication made use  of data products from the  Two Micron All Sky Survey,  which is a joint  project of the University of  Massachusetts and  the Infrared Processing  and Analysis Center/California  Institute of  Technology,  funded by  NASA and  the National Science Foundation.
  
{\it Facilities:} \facility{Hale (PALM-3000, PHARO)},  \facility{Gemini (Hokupa\`{}a, QUIRC)} 
 


\end{document}